\documentclass[prd,reprint,amsmath,amssymb,superscriptaddress]{revtex4-2}

\usepackage{graphicx}
\usepackage[colorlinks=true,linkcolor=blue,citecolor=blue,urlcolor=blue]{hyperref}

\begin{document}

\title{Fast Population Leakage in Astronomical Masers:
Maser Amplification and Transient Superradiance}

\author{Vahid Anari}
\affiliation{Department of Physics and Astronomy, McMaster University, 1280 Main Street West, Hamilton, Ontario L8S 4M1, Canada}
\affiliation{Department of Physics and Astronomy, The University of Western Ontario, 1151 Richmond Street, London, Ontario N6A 3K7, Canada}
\author{Toktam Rashidi}
\affiliation{Department of Physics and Astronomy, McMaster University, 1280 Main Street West, Hamilton, Ontario L8S 4M1, Canada}
\author{Fereshteh Rajabi}
\email{rajabf1@mcmaster.ca}
\affiliation{Department of Physics and Astronomy, McMaster University, 1280 Main Street West, Hamilton, Ontario L8S 4M1, Canada}

\begin{abstract}

We use a \(\Lambda\)-type three-level Maxwell--Bloch model to test whether an inverted molecular transition in an astronomical maser source can produce maser amplification or superradiance when its upper level also decays through a second radiative pathway with a much larger spontaneous decay rate. Such shared-upper-level configurations occur in multilevel, radiatively pumped molecules, including Class~II methanol masers and several OH maser transitions. The model follows the coupled evolution of level populations, molecular coherences, radiation fields, and phenomenological relaxation and dephasing, separating population leakage from coherence loss. We focus on the mixed configuration in which the observed transition is inverted while the faster pathway is non-inverted and acts as a leakage channel. We find that rapid spontaneous decay through the competing pathway does not, by itself, suppress maser amplification or superradiant emission from the inverted transition. The response is controlled by the shared upper-level population reservoir, the available initial coherence, and the relaxation and dephasing timescales. For small effective coherence in the leakage pathway, its radiative output remains weak, while the inverted transition either amplifies a seed field in the quasi-steady maser regime or develops macroscopic coherence and produces a transient superradiant burst. A larger inversion does not necessarily produce a stronger burst if it is accompanied by a weaker initial coherence seed. As a benchmark, we apply the model to the 6.7~GHz methanol flare in S255IR-NIRS3, whose upper level also decays through the 239.7~GHz transition at a spontaneous rate more than four orders of magnitude larger. The calculated flare remains compatible with a transient-superradiance interpretation when this fast leakage pathway is included explicitly.

\end{abstract}


\maketitle

\section{Introduction}
\label{sec:Introduction}

Astronomical masers are sensitive probes of the physical and kinematic conditions in star-forming regions and other molecular environments. Their intense, spectrally narrow emission and strong variability make them powerful diagnostics of gas density, temperature, and velocity structure, as well as useful tools for measuring distances and magnetic field strengths \citep{Weaver1965, Ellingsen2006}. Classical maser theory often treats a masing transition as an effective two-level system, with excitation and decay described by rate equations for the level populations together with radiative transfer for the emergent intensity \citep{Elitzur1992, Gray2012}. Although this approach has successfully explained many observed masing phenomena, several sources display temporal behaviour that is difficult to reproduce within this framework, including rapid flux-density increases during flares and pump--flare variations in which the maser emission does not simply track changes in the pump \citep{Araya2010, VanderWalt2011, Gray2018, Szymczak2018a, Szymczak2018b, Goedhart2019, VandenHeever2019}. Such behaviours suggest the presence of additional physical processes not captured by classical maser theory.

Motivated by Dicke's theory of cooperative spontaneous emission \citep{Dicke1954}, superradiance has been introduced in astrophysical maser studies as a transient cooperative emission regime distinct from quasi-steady maser action \citep{Rajabi2016B, Rajabi2017, Rajabi2019}. In a unified Maxwell--Bloch framework, masers and superradiance emerge as different dynamical regimes: superradiance corresponds to rapid cooperative emission enabled by the buildup of macroscopic molecular coherence, whereas classical maser action corresponds to stimulated amplification in a quasi-steady regime where the molecular coherence remains comparatively weak \citep{Feld1980, Rajabi2020, Houde2024}. The Maxwell--Bloch equations (MBEs) have reproduced variability in several maser-hosting sources, including 6.7~GHz and 12.2~GHz Class~II methanol flares \citep{Rajabi2019, Rajabi2023, Rashidi2025, Rashidi2026}, OH 18~cm flares \citep{Rajabi2016B, Rajabi2023}, and 22~GHz water flares \citep{Rajabi2017}.

In all of these applications, the observed transition has been treated as an effective two-level subsystem, with the surrounding molecular network and environmental interactions represented through effective pumping, relaxation, and dephasing terms. This approximation is useful, but by construction it absorbs rather than explicitly follows additional radiative pathways that share the same upper level as the observed transition. In some multilevel molecules, particular maser transitions have shared-upper-level radiative pathways with spontaneous decay rates far exceeding that of the observed maser line. Examples include Class~II methanol transitions at 6.7~GHz, 12.2~GHz, and 107~GHz, and OH transitions at 4765~MHz, 6030~MHz, and 6035~MHz \citep{CDMS2001}.

These fast shared-upper-level pathways motivate the central question addressed in this work: whether rapid population loss through a competing transition can suppress maser amplification or superradiance, and under what conditions sufficient population inversion can survive to sustain maser amplification or, together with the buildup of macroscopic coherence, produce superradiant emission. To address this question, we solve the three-level MBEs for the \(\Lambda\)-type system shown in Fig.~\ref{fig:three-level-diagram}, where the upper state \(\lvert 3\rangle\) is shared by two radiative transitions. Throughout the paper, \(3\rightarrow 2\) denotes the observed inverted transition, while \(3\rightarrow 1\) denotes the fast shared-upper-level leakage pathway.

The same equations can be applied to different initial population configurations. Here we focus on the mixed configuration relevant to shared-upper-level astronomical masers: the leakage pathway is non-inverted, while the observed transition is inverted. In this configuration, the effect of the fast pathway is not set by its spontaneous decay rate alone; it also depends on the available population reservoir, the coherence dynamics in each transition, and the non-coherent relaxation and dephasing processes.

We first examine generic finite-inversion cases within this mixed configuration to map the competition between leakage and coherence buildup. We then apply the same framework to the 6.7~GHz methanol flare in S255IR-NIRS3, which represents the strong-inversion end of this configuration and includes a known fast 239.7~GHz shared-upper-level decay pathway. This source provides a realistic benchmark for testing whether the previously proposed superradiance interpretation \citep{Rajabi2019} remains viable when the fast leakage pathway is included explicitly.

The paper is organized as follows. 
Section~\ref{sec:modeloverview} presents the three-level 
Maxwell--Bloch model and introduces the initial and boundary 
conditions for the \(\Lambda\)-type system. The assumptions leading 
to the equations and the scaling adopted in this work are summarized 
in \hyperref[app:three_level_mbes]{Appendix~A}. 
Section~\ref{sec:analysis} investigates the competition between fast 
leakage and coherence buildup through numerical solutions of the 
three-level equations, with particular attention to the roles of the 
population inversion, coherence development, and non-coherent 
relaxation and dephasing. That section also includes the application 
of the model to the 6.7~GHz methanol flare in S255IR-NIRS3. 
Supporting scans of the non-coherent relaxation and dephasing 
timescales in the non-inverted leakage pathway are provided in 
\hyperref[app:noncoherent_effects]{Appendix~B}. 
Section~\ref{sec:conclusion} summarizes the physical implications for 
radiatively pumped maser and superradiance systems with fast competing 
decay pathways.

\section{Three-level Maxwell--Bloch model}
\label{sec:modeloverview}

\begin{figure}[t]
    \centering
    \includegraphics[width=0.7\linewidth]{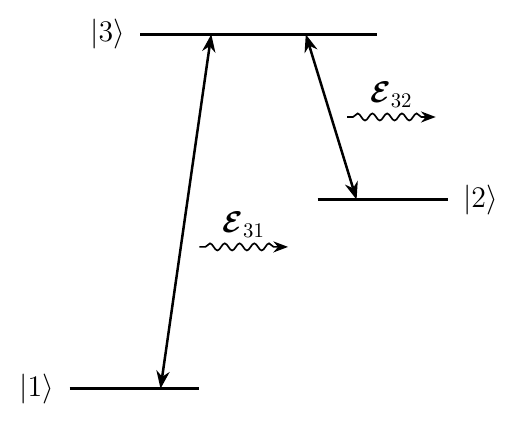}
    \caption{Level notation for the \(\Lambda\)-type three-level model. The transitions \(3\rightarrow1\) and \(3\rightarrow2\) are dipole allowed and are coupled to the field envelopes \(E_{31}\) and \(E_{32}\), respectively, while the \(2\leftrightarrow1\) transition is taken to be dipole forbidden. The level spacings are schematic and not to scale.}
    \label{fig:three-level-diagram}
\end{figure}

We specify the three-level Maxwell--Bloch model associated with the notation in Fig.~\ref{fig:three-level-diagram}. The upper state \(|3\rangle\) is shared by the two dipole-allowed transitions \(3\rightarrow1\) and \(3\rightarrow2\), which are coupled to the field envelopes \(E_{31}\) and \(E_{32}\), respectively, while \(2\leftrightarrow1\) is taken to be dipole forbidden. In the astrophysical interpretation introduced above, \(3\rightarrow2\) is the observed inverted transition and \(3\rightarrow1\) is the fast shared-upper-level leakage pathway; the equations in this section are written generally, and the relevant population regime is specified through the initial conditions.

We consider an ensemble of these three-level molecules with number density \(n_0\) in a cylindrical sample of length \(L\) and cross-sectional area \(A\), with the cylinder axis taken as the \(z\)-direction. Since the observed emission, whether produced by maser amplification or superradiance, is associated with the \(3\rightarrow 2\) transition, we define the Fresnel number using its wavelength \(\lambda_{32}\),
\[
F_{32}=\frac{A}{\lambda_{32}L},
\]
and take \(F_{32}\simeq 1\). This choice makes a one-dimensional Maxwell--Bloch model a good approximation for describing the matter--field dynamics and radiation propagation on the \(3\rightarrow 2\) transition, while the \(3\rightarrow 1\) field is treated as a competing radiative channel propagating through the same column. The equations are written in the retarded time \(t_{\rm ret}=t-z/c\) and are obtained under the rotating-wave and slowly varying envelope approximations for resonant fields \citep{Gross1982, Benedict1996}. A compact derivation and the corresponding unscaled equations are given in \hyperref[app:three_level_mbes]{Appendix~A}.

The medium is described by normalized level populations \(n_i\), so that the physical number density in level \(i\) is \(n_0 n_i\), and by slowly varying coherence variables \(R_{ij}\) associated with the \(i\leftrightarrow j\) transition pairs. For the dipole-allowed transitions, \(R_{31}\) and \(R_{32}\), together with the corresponding dipole moments and number density, determine the polarization envelopes that source the fields \(E_{31}\) and \(E_{32}\). The lower-state coherence \(R_{21}\) is retained because it can be generated indirectly through the two allowed transitions, although no field is applied directly to the dipole-forbidden \(2\leftrightarrow1\) transition. We also define population differences \(n_{3j}\equiv n_3-n_j\), so that positive \(n_{3j}\) corresponds to inversion on the \(3\rightarrow j\) transition.

Non-coherent processes are included phenomenologically. We use \(T_{3j}^{(1)}\), with \(j=1,2\), to denote the relaxation timescales of the population differences \(n_{3j}\), and \(T_{ij}^{(2)}\), with \(ij=31,32,21\), to denote the coherence dephasing timescales of the corresponding \(i\leftrightarrow j\)  pairs. For the dipole-allowed transitions, this dephasing damps the polarization envelopes that source the radiation fields. These timescales enter in the relaxation and dephasing terms in the equations below. In a collisional medium, coherence dephasing is often faster than population relaxation; we therefore take \(T_{3j}^{(1)}>T_{3j}^{(2)}\) for \(j=1,2\), because elastic collisions can randomize molecular phases without changing level populations, while population-changing collisions also interrupt phase coherence \citep{Gross1982,Benedict1996}.

For the numerical calculations, we use the scaled variables
\[
\xi=\frac{z}{L}, \qquad \tau=\frac{t_{\rm ret}}{T_0},
\]
where \(L\) is the sample length and \(T_0\) is a reference timescale
that is so far arbitrary. The scaled field envelopes and non-coherent
timescales are
\begin{equation}
\tilde{E}_{3j}=-i\,\frac{T_0d_{3j}E_{3j}}{\hbar},
\label{eq:scaled_field}
\end{equation}
and
\begin{equation}
\tau_{ij}^{(k)}=\frac{T_{ij}^{(k)}}{T_0},
\label{eq:scaled_times}
\end{equation}
where \(k=1,2\).

With these definitions, the matter (Bloch) Eqs.~\eqref{eq:app_matter} in scaled form are
\begin{subequations}
\label{eq:matter_dimless}
\begin{align}
\frac{\partial n_{31}}{\partial\tau} &= 
   -\operatorname{Re}\!\left(\tilde{E}_{31}^{*}R_{31}
   +\frac{1}{2}\tilde{E}_{32}^{*}R_{32}\right)
   -\frac{n_{31}-n^{0}_{31}}{\tau^{(1)}_{31}}, 
   \label{eq:n31_dimless}\\[3pt]
\frac{\partial n_{32}}{\partial\tau} &= 
   -\operatorname{Re}\!\left(\frac{1}{2}\tilde{E}_{31}^{*}R_{31}
   +\tilde{E}_{32}^{*}R_{32}\right)
   -\frac{n_{32}-n^{0}_{32}}{\tau^{(1)}_{32}}, 
   \label{eq:n32_dimless}\\[3pt]
\frac{\partial R_{31}}{\partial\tau} &= 
   \tilde{E}_{31} n_{31}
   - \frac{1}{2}\tilde{E}_{32} R_{21}
   - \frac{R_{31}-R^{0}_{31}}{\tau^{(2)}_{31}}, 
   \label{eq:R31_dimless}\\[3pt]
\frac{\partial R_{32}}{\partial\tau} &= 
   \tilde{E}_{32} n_{32}
   - \frac{1}{2}\tilde{E}_{31} R^{*}_{21}
   - \frac{R_{32}-R^{0}_{32}}{\tau^{(2)}_{32}}, 
   \label{eq:R32_dimless}\\[3pt]
\frac{\partial R_{21}}{\partial\tau} &= 
   \frac{1}{2}\!\left(\tilde{E}_{31}R^{*}_{32}
   +\tilde{E}_{32}^{*}R_{31}\right)
   - \frac{R_{21}-R^{0}_{21}}{\tau^{(2)}_{21}},
   \label{eq:R21_dimless}
\end{align}
\end{subequations}
and the corresponding field (Maxwell) Eqs.~\eqref{eq:app_E3j} for the scaled field
envelopes are
\begin{equation}
\frac{\partial \tilde{E}_{3j}}{\partial \xi}
=\frac{n_0Lk_{3j}|d_{3j}|^2}{2\epsilon_0\hbar}\,T_0\,R_{3j},
\label{eq:E3j_dimless}
\end{equation}
where \(\operatorname{Re}(\cdot)\) denotes the real part of the entire
enclosed expression, an asterisk denotes complex conjugation, and
\(d_{3j}\) and \(k_{3j}=\omega_{3j}/c\) denote the transition dipole
moment and wavenumber, respectively, of the \(3\rightarrow j\)
transition, with \(\omega_{3j}\) the corresponding angular transition
frequency and \(c\) the speed of light. Choosing
\begin{equation}
T_0=\frac{2\epsilon_0\hbar}{n_0Lk_{32}|d_{32}|^2},
\label{eq:T0}
\end{equation}
normalizes the field coupling of the inverted \(3\rightarrow2\)
transition to unity, and Eq.~\eqref{eq:E3j_dimless} becomes
\begin{subequations}
\label{eq:field_dimless}
\begin{align}
\frac{\partial \tilde{E}_{31}}{\partial \xi} &= \eta\, R_{31},
\label{eq:E1_dimless}\\[3pt]
\frac{\partial \tilde{E}_{32}}{\partial \xi} &= R_{32},
\label{eq:E2_dimless}
\end{align}
\end{subequations}
where
\begin{equation}
\eta=\left|\frac{d_{31}}{d_{32}}\right|^2\frac{k_{31}}{k_{32}}
\label{eq:eta}
\end{equation}
is the relative field-coupling strength of the \(3\rightarrow1\) pathway
compared with the \(3\rightarrow2\) pathway. Thus, once the two molecular
transitions are specified, \(\eta\) is fixed by their spectroscopic
properties and is not an additional fitting parameter.

In the population equations, \(n_{31}^{0}\) and \(n_{32}^{0}\) specify the reference population differences toward which the phenomenological relaxation terms drive the medium in the absence of matter--field exchange. In particular, the relaxation term may be written as
\[
-\frac{n_{3j}-n_{3j}^{0}}{\tau_{3j}^{(1)}}
=
-\frac{n_{3j}}{\tau_{3j}^{(1)}}
+
\frac{n_{3j}^{0}}{\tau_{3j}^{(1)}},
\]
where the first contribution relaxes the population difference, while the second acts as an effective steady source that maintains the reference value \(n_{3j}^{0}\). Thus, when finite relaxation is included, \(n_{31}^{0}\) and \(n_{32}^{0}\) represent the background population state established by steady pumping and non-coherent population-transfer processes. To close Eqs.~\eqref{eq:matter_dimless} and \eqref{eq:field_dimless}, we must specify the initial populations, initial coherences, field boundary conditions, and non-coherent relaxation and dephasing timescales. We begin with the initial populations.

The initial normalized populations satisfy
\[
0\le n_i^0\le1,\qquad \sum_{i=1}^{3}n_i^0=1.
\]
We parameterize them using the ratios
\[
r_{3j}=\frac{n_3^0}{n_j^0},\qquad j=1,2.
\]
The corresponding normalized populations are
\begin{equation}
\label{eq:initial_n}
\begin{split}
n_{1}^{0} &= \frac{r_{32}}{\,r_{31}+r_{32}+r_{31}r_{32}\,},\\
n_{2}^{0} &= \frac{r_{31}}{\,r_{31}+r_{32}+r_{31}r_{32}\,},\\
n_{3}^{0} &= \frac{r_{31}r_{32}}{\,r_{31}+r_{32}+r_{31}r_{32}\,}.
\end{split}
\end{equation}
This parameterization automatically satisfies the normalization and positivity constraints on the initial populations \(n_i^0\) for positive \(r_{31}\) and \(r_{32}\). The corresponding reference population differences are taken to coincide with the spatially uniform initial population differences:
\begin{equation}
\label{eq:population_inversion}
\begin{split}
n_{31}^{0}
\equiv n_{31}(\xi,\tau=0)
&= \frac{r_{32}(r_{31}-1)}
 {\,r_{31}+r_{32}+r_{31}r_{32}\,},\\
n_{32}^{0}
\equiv n_{32}(\xi,\tau=0)
&= \frac{r_{31}(r_{32}-1)}
 {\,r_{31}+r_{32}+r_{31}r_{32}\,}.
\end{split}
\end{equation}
For positive \(r_{31}\) and \(r_{32}\), the denominator in Eq.~\eqref{eq:population_inversion} is positive. The signs of the initial population differences are therefore determined by \(r_{31}-1\) and \(r_{32}-1\). The mixed configuration analyzed here corresponds to \(r_{31}<1,\ r_{32}>1\): the \(3\rightarrow1\) pathway is non-inverted and is identified with the leakage channel, while \(3\rightarrow2\) is the inverted maser/superradiant transition. Other choices of \(r_{31}\) and \(r_{32}\) correspond to the remaining population configurations and can be treated within the same formalism.

At \(\tau=0\), we denote the initial coherence amplitudes by
\begin{equation}
    R_{ij}^0\equiv R_{ij}(\xi, 0).
\end{equation}
For a physical initial state, the magnitude of each coherence is bounded by the corresponding level populations through the Cauchy--Schwarz inequality. With the normalization used here,
\begin{equation}
|R^0_{ij}|\le \mathcal{R}_{ij}^{\max}
=2\sqrt{n_i^0n_j^0}.
\label{eq:Rmax}
\end{equation}
We write
\[
R^0_{ij}=c_{ij}\mathcal{R}_{ij}^{\max},
\]
where \(|c_{ij}|\le1\). The parameter \(c_{ij}\) is the initial coherence fraction for the \(i\leftrightarrow j\) pair. Since the \(2\leftrightarrow1\) transition is dipole forbidden and no direct field is applied to it, we set \(c_{21}=0\). Any lower-state coherence \(R_{21}\) is therefore generated only indirectly through the two allowed transitions. The initial state of the system is therefore specified by
\[
\{r_{31},r_{32},c_{31},c_{32}\}.
\]
No incident radiation is imposed at the entrance of the sample, and the calculated emission is generated by the medium itself. Thus, the boundary conditions are
\begin{equation}
\tilde{E}_{3j}(\xi=0,\tau)=0.
\label{eq:BCs}
\end{equation}

For spatially uniform initial coherences, Eqs.~\eqref{eq:field_dimless} can be integrated directly at \(\tau=0\). At the output end of the sample, \(\xi=1\), this gives
\begin{equation}
\tilde{E}_{31}(1,0)=\eta\,R^0_{31},\qquad
\tilde{E}_{32}(1,0)=R^0_{32}.
\label{eq:initial_fields}
\end{equation}
The corresponding scaled initial intensities are therefore proportional to
\begin{equation}
\tilde{I}_{31}(1,0)\propto
\eta^2|R^0_{31}|^2,
\qquad
\tilde{I}_{32}(1,0)\propto
|R^0_{32}|^2.
\label{eq:initial_intensity}
\end{equation}

Together with these initial and boundary conditions, Eqs.~\eqref{eq:matter_dimless} and \eqref{eq:field_dimless} define the three-level Maxwell--Bloch model analyzed in Sec.~\ref{sec:analysis}. These coupled field-matter equations are solved numerically using a fourth-order Runge--Kutta scheme. 


\section{Competition Between Fast Leakage and Coherence Buildup}
\label{sec:analysis}

We restrict the numerical study to the mixed population configuration
relevant to the shared-upper-level Class~II methanol and OH systems
discussed above:
\begin{equation}
n_3^0<n_1^0,\quad n_3^0>n_2^0
\quad\Longleftrightarrow\quad
r_{31}<1,\quad r_{32}>1 .
\label{eq:mixed_regime}
\end{equation}
The observed \(3\rightarrow2\) transition is taken to be inverted and
can therefore support either quasi-steady maser amplification or superradiance. In contrast, no maser
or superradiant emission has been reported from the faster
\(3\rightarrow1\) pathway, motivating its treatment as a non-inverted
leakage channel. This population ordering is consistent with an
approximately thermalized \(3\rightarrow1\) subsystem, although whether that
subsystem remains dynamically thermalized depends on the relaxation and
dephasing timescales adopted in each calculation. The coexistence of an
inverted transition and non-inverted, reservoir-coupled transitions
within the same three-level system is familiar from three-level maser
models \citep{Scovil1959,Geva1994}.

The following subsections isolate the three ingredients that determine
whether the \(3\rightarrow2\) transition can sustain maser amplification
or superradiant emission despite fast leakage: the shared-upper-level
population reservoir, the initial coherence in each radiative channel,
and the non-coherent relaxation and dephasing timescales. In Secs.~\ref{subsec:InversionANDcollective} and
\ref{subsec:coherence_levels}, all non-coherent timescales are set to
infinity, so that population restoration and coherence damping are
absent. This allows the effects of the initial population distribution
and coherence fractions to be examined independently of non-coherent
processes. When finite relaxation and dephasing timescales are introduced in
Sec.~\ref{subsec:noncoherent_regimes}, we adopt the hierarchy
\begin{equation}
\label{eq:timescales_inequality}
\tau_{31}^{(k)} \lesssim \tau_{32}^{(k)}, \qquad k=1,2 ,
\end{equation}
where \(k=1\) denotes population relaxation and \(k=2\) denotes loss of phase coherence. This ordering represents the assumption that non-coherent processes act at least as efficiently in the approximately thermalized \(3\rightarrow1\) leakage pathway as in the inverted \(3\rightarrow2\) transition.

The initial coherences \(R^0_{31}\) and
\(R^0_{32}\) are treated independently of the timescale
hierarchy. In Sec.~\ref{subsec:coherence_levels}, we vary the
corresponding coherence fractions \(c_{31}\) and \(c_{32}\) to determine
how the initial coherence in each radiative pathway affects its output.
Throughout Secs.~\ref{subsec:InversionANDcollective}--%
\ref{subsec:noncoherent_regimes}, we set \(\eta=1\), so that the
\(3\rightarrow1\) and \(3\rightarrow2\) transitions have equal scaled
field-coupling strengths. By removing the asymmetry arising from their
spectroscopic coupling strengths, this choice allows differences between
the two radiative channels to be attributed to the initial population
distribution, coherence fractions, and non-coherent relaxation and
dephasing timescales. For the S255IR-NIRS3 methanol application
in Sec.~\ref{subsec:S255}, we use the spectroscopic value
\(\eta=27.51\), computed from the dipole moments and wavenumbers of the
239.7~GHz \(3\rightarrow1\) leakage transition and the 6.7~GHz
\(3\rightarrow2\) methanol transition.


\subsection{Population Reservoir and Collective Emission}
\label{subsec:InversionANDcollective}

\begin{figure*}
    \centering
    \begin{minipage}{0.49\textwidth}
        \centering
        \textbf{(a)}\\[-1mm]
        \includegraphics[width=\linewidth]{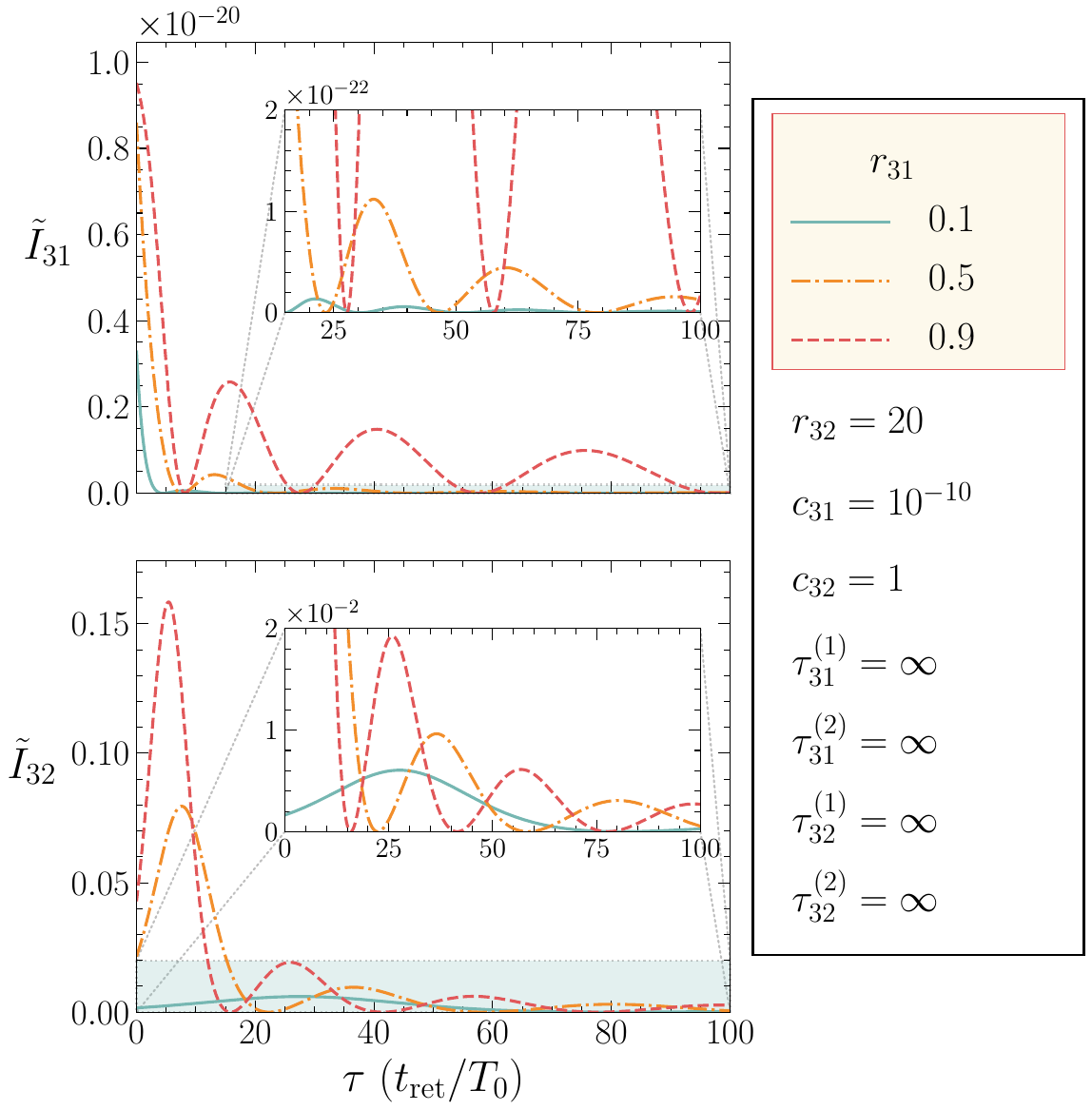}
    \end{minipage}
    \hfill
    \begin{minipage}{0.49\textwidth}
        \centering
        \textbf{(b)}\\[-1mm]
        \includegraphics[width=\linewidth]{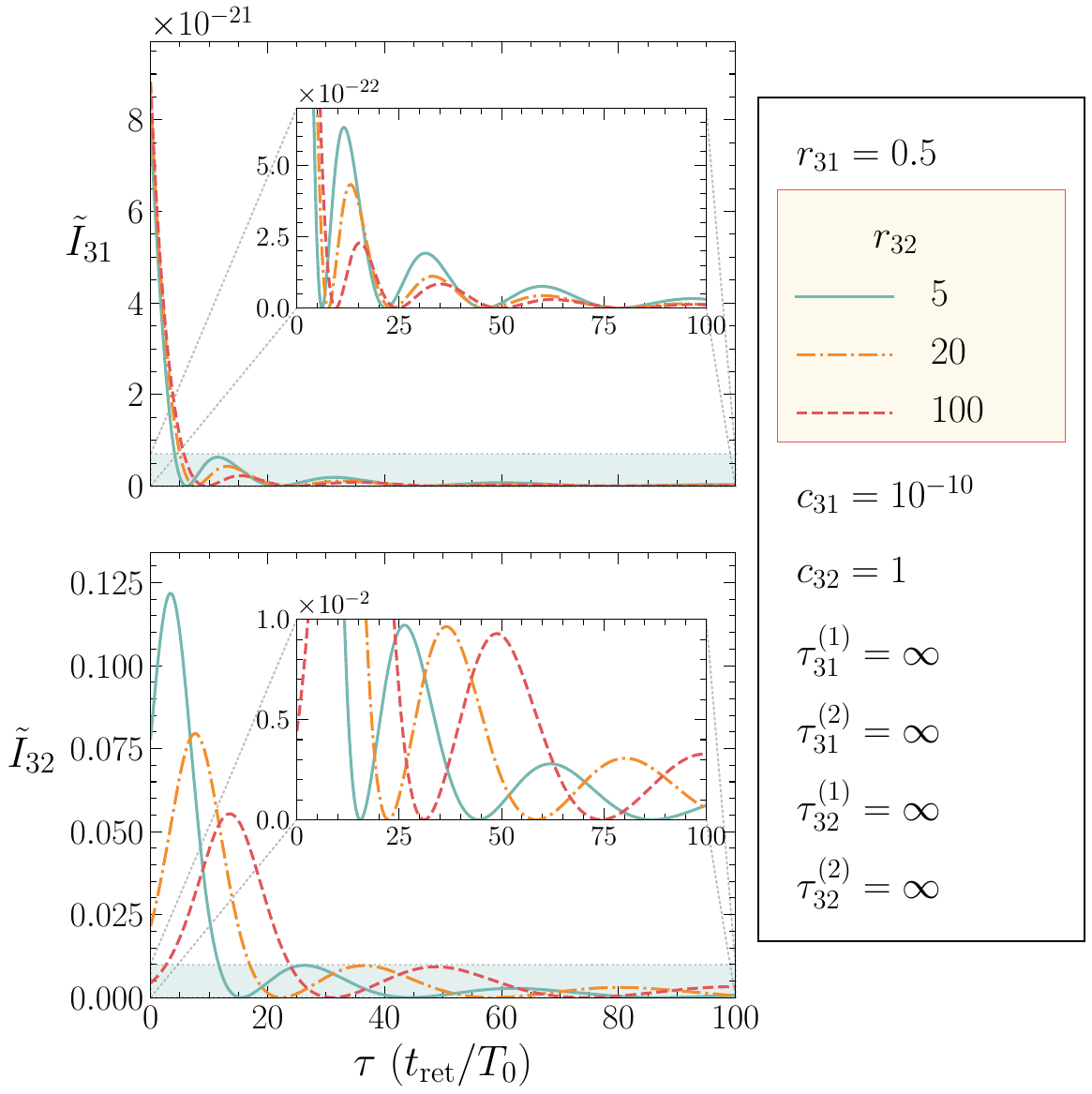}
    \end{minipage}
\caption{Population-ratio dependence of the scaled output intensities
in the \(\Lambda\)-type three-level model. The scaled intensities
\(\tilde{I}_{31}(1,\tau)\) and \(\tilde{I}_{32}(1,\tau)\) are shown as
functions of the scaled retarded time \(\tau\) at the output end of the
sample, \(\xi=1\). Panel (a) varies \(r_{31}=n_3^0/n_1^0\) at fixed
\(r_{32}=20\). Panel (b) varies \(r_{32}=n_3^0/n_2^0\) at fixed \(r_{31}=0.5\). In both panels,
\(\eta=1\), \(c_{31}=10^{-10}\), \(c_{32}=1\), and all non-coherent
relaxation and dephasing timescales are set to infinity, so the figure
isolates the effect of the initial population distribution in an
idealized transient-superradiance limit. The different vertical scales
emphasize the asymmetry between the weak coherent output from the
\(3\rightarrow1\) leakage channel and the collective emission from the
inverted \(3\rightarrow2\) subsystem.}
\label{fig:case_population}
\end{figure*}

We first investigate the role of the initial population distribution.
As noted above, all non-coherent relaxation and dephasing timescales in
Fig.~\ref{fig:case_population} are set to infinity. Consequently,
neither population relaxation nor coherence damping acts during these
calculations, and the \(3\rightarrow1\) subsystem is not dynamically
thermalized. Instead, the non-inverted leakage pathway is initialized
with the small coherence fraction \(c_{31}=10^{-10}\), representing the
negligible initial polarization expected for an approximately
thermalized transition. This choice specifies a nearly incoherent
initial state but does not replace the thermalizing effects of finite
relaxation and dephasing. On the other hand, the inverted \(3\rightarrow2\) transition is strongly seeded with \(c_{32}=1\). For this transition, the absence of dephasing permits macroscopic phase coherence to develop, while the absence of population relaxation prevents sustained replenishment of the inversion. The resulting response therefore lies in the transient superradiance regime rather than the quasi-steady maser regime \citep{Rajabi2020}.

Using Eqs.~\eqref{eq:Rmax} and \eqref{eq:initial_intensity}, the initial
scaled output intensities at the end of the sample, \(\xi=1\), are,
up to fixed factors involving \(\eta\) and the coherence fractions,
proportional to
\begin{equation}
\label{eq:initial_I_exact}
\begin{aligned}
    \tilde{I}_{31}(1,\tau=0) &\propto
    \frac{r_{31} r_{32}^2}{(r_{31}+r_{32}+r_{31}r_{32})^2},\\
    \tilde{I}_{32}(1,\tau=0) &\propto
    \frac{r_{31}^2 r_{32}}{(r_{31}+r_{32}+r_{31}r_{32})^2}.
\end{aligned}
\end{equation}
The corresponding leading initial slopes, neglecting non-coherent terms, scale as
\begin{equation}
\label{eq:initial_slope_exact}
\begin{aligned}
     \frac{\partial \tilde{I}_{31}}{\partial \tau}
     &\propto
     \frac{r_{31} r_{32}^3(r_{31}-1)}{(r_{31}+r_{32}+r_{31}r_{32})^3},\\
    \frac{\partial \tilde{I}_{32}}{\partial \tau}
    &\propto
    \frac{r_{31}^3 r_{32}(r_{32}-1)}{(r_{31}+r_{32}+r_{31}r_{32})^3}.
\end{aligned}
\end{equation}
The signs of these slopes reflect the contrasting early-time responses
of the two transitions in the mixed configuration. Since \(r_{31}<1\),
the intensity in the non-inverted \(3\rightarrow1\) channel initially
declines, whereas \(r_{32}>1\) causes the intensity in the inverted
\(3\rightarrow2\) channel to grow. In the absence of dephasing, the
initial \(3\rightarrow2\) coherence seeds the cooperative buildup of
macroscopic polarization, ultimately producing a superradiant burst.

Figure~\ref{fig:case_population}(a) shows the effect of increasing
\(r_{31}\) at fixed \(r_{32}\). Larger \(r_{31}\) places more population in the
shared upper level \(\lvert 3\rangle\), increasing the reservoir
available to the inverted \(3\rightarrow2\) transition. As a result,
the \(\tilde{I}_{32}\) burst occurs earlier and reaches a higher peak
intensity. In the absence of non-coherent processes, the time of the
first intensity maximum corresponds to the superradiant delay time,
during which macroscopic polarization develops before the burst
\citep{Gross1982,Rajabi2020}.

Figure~\ref{fig:case_population}(b) shows the complementary effect of increasing \(r_{32}\) at fixed \(r_{31}\). Larger \(r_{32}\) lowers the initial population in level 2 and therefore increases the inversion \(n_{32}^0\). However, the maximum allowed initial coherence on the inverted transition depends on the product \(n_3^0 n_2^0\), not only on the inversion. From Eq.~\eqref{eq:Rmax},
\begin{equation}
\mathcal{R}_{32}^{\max}
=
2\sqrt{n_3^0n_2^0}
=
\frac{2r_{31}\sqrt{r_{32}}}{r_{31}+r_{32}+r_{31}r_{32}}.
\end{equation}
Thus, although increasing \(r_{32}\) strengthens the population inversion, it reduces the coherence seed available on the \(3\rightarrow2\) transition. For fixed \(c_{32}\), this explains why the \(\tilde I_{32}\) burst in Fig.~\ref{fig:case_population}(b) becomes weaker and appears at later times as \(r_{32}\) increases. The longer delay is consistent with superradiant buildup from a weaker initial coherence seed.

The finite-\(r_{32}\) calculations in Fig.~\ref{fig:case_population}(b) also show the trend toward the strong-inversion limit considered later for the 6.7~GHz methanol flare in S255IR-NIRS3. For \(r_{31}<1\) and \(r_{32}\gg1\), \(r_{31}+r_{32}+r_{31}r_{32}\simeq r_{32}(1+r_{31})\), so the maximum initial coherence on the inverted transition becomes
\begin{equation}
\mathcal{R}_{32}^{\max}\simeq
\frac{2r_{31}}{(1+r_{31})\sqrt{r_{32}}} .
\end{equation}
Thus, for fixed \(c_{32}\), the strong-inversion limit is also an
increasingly weak-coherence-seed limit, since the initial
\(3\rightarrow2\) coherence decreases as \(r_{32}^{-1/2}\). The corresponding large-\(r_{32}\) initial intensity scalings at \(\xi=1\) are
\[
\begin{aligned}
\tilde{I}_{31}(1,\tau=0)&\propto \frac{r_{31}}{(1+r_{31})^2},\\
\tilde{I}_{32}(1,\tau=0)&\propto \frac{1}{r_{32}}
\left(\frac{r_{31}}{1+r_{31}}\right)^2.
\end{aligned}
\]
These limits show that the initial leakage-channel intensity saturates as \(r_{32}\) grows, whereas the initial intensity on the inverted transition continues to decrease as \(1/r_{32}\).

Figure~\ref{fig:case_population} demonstrates that the collective
\(3\rightarrow2\) response is controlled by both the shared
upper-level population reservoir and the available initial coherence,
rather than by population inversion alone. Increasing \(r_{31}\) raises
the population stored in \(\lvert 3\rangle\), producing an earlier and
stronger burst. Increasing \(r_{32}\) at fixed \(r_{31}\) strengthens the
inversion \(n_{32}^{0}\), but lowers \(n_2^{0}\) and therefore reduces
\(\mathcal{R}_{32}^{\max}\), delaying and weakening the burst. For the
adopted small leakage-channel coherence fraction
\(c_{31}=10^{-10}\), the \(3\rightarrow1\) intensity remains many
orders of magnitude below the collective \(3\rightarrow2\) output
throughout both panels, as indicated by their different vertical
scales.


\subsection{Coherence Seeding and Collective Emission}
\label{subsec:coherence_levels}

\begin{figure*}
    \centering
    \begin{minipage}{0.49\textwidth}
        \centering
        \textbf{(a)}\\[-1mm]
        \includegraphics[width=\linewidth]{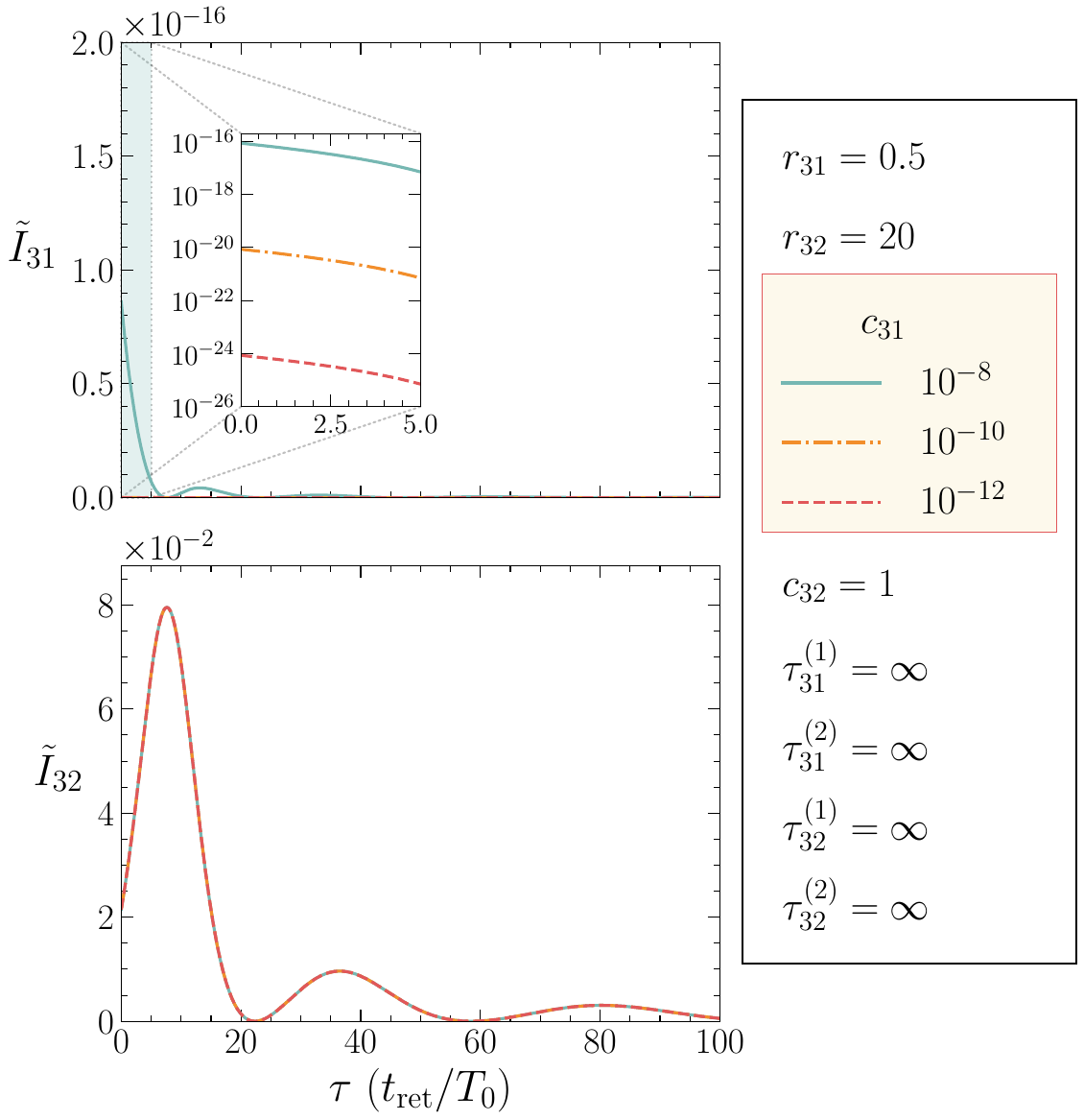}
    \end{minipage}
    \hfill
    \begin{minipage}{0.49\textwidth}
        \centering
        \textbf{(b)}\\[-1mm]
        \includegraphics[width=\linewidth]{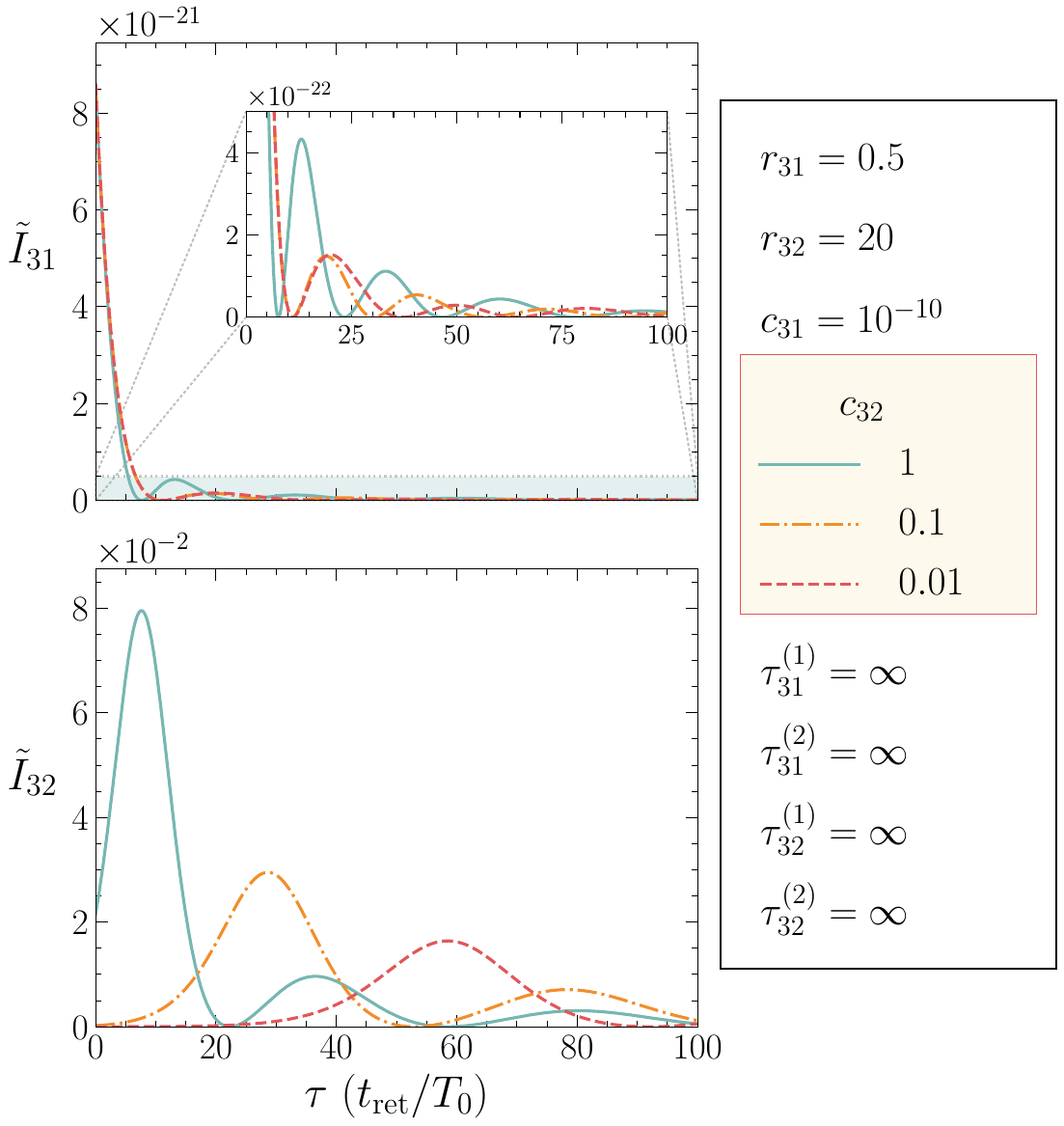}
    \end{minipage}
    \caption{Initial-coherence dependence of the scaled output
    intensities in the two radiative channels. The intensities
    \(\tilde{I}_{31}(1,\tau)\) and \(\tilde{I}_{32}(1,\tau)\) are shown
    at the output end of the sample, \(\xi=1\), as functions of the
    scaled retarded time \(\tau\). Panel (a) varies \(c_{31}\), and
    hence the initial coherence of the non-inverted
    \(3\rightarrow1\) leakage pathway, at fixed \(c_{32}=1\). Panel
    (b) varies \(c_{32}\), and hence the initial coherence of the
    inverted \(3\rightarrow2\) transition, at fixed
    \(c_{31}=10^{-10}\). In both panels, \(\eta=1\), \(r_{31}=0.5\),
    \(r_{32}=20\), and all non-coherent relaxation and dephasing
    timescales are set to infinity, so that the effects of coherence
    seeding are examined in the transient-superradiance limit.
    Increasing \(c_{31}\) strengthens the weak leakage-channel
    emission, whereas increasing \(c_{32}\) shortens the delay time
    and strengthens the cooperative burst.}
    \label{fig:case_coherence}
\end{figure*}

In this subsection, we examine the role of the initial coherences.
Throughout the calculations, the population ratios are fixed at
\(r_{31}=0.5\) and \(r_{32}=20\), and all non-coherent relaxation and
dephasing timescales are set to infinity. With the population
distribution fixed, the initial coherence on each transition is
specified through
\(R^0_{ij}=c_{ij}\mathcal{R}_{ij}^{\max}\). We therefore vary
\(c_{31}\) and \(c_{32}\) separately, holding the other fixed, to
distinguish coherence-seeding effects from the population-reservoir
effects examined in the preceding subsection.

From Eq.~\eqref{eq:initial_intensity}, the corresponding initial scaled
intensities satisfy
\[
\tilde{I}_{31}(1,0)\propto c_{31}^{\,2},
\qquad
\tilde{I}_{32}(1,0)\propto c_{32}^{\,2},
\]
up to fixed factors involving \(\eta\) and
\(\mathcal{R}_{ij}^{\max}\). Although both intensities depend
quadratically on their respective coherence fractions, their subsequent
dynamics differ because only the \(3\rightarrow2\) transition is
inverted.

Figure~\ref{fig:case_coherence}(a) shows that increasing \(c_{31}\)
enhances the \(3\rightarrow1\) output. For the values considered,
however, this emission remains weak compared with the cooperative
\(3\rightarrow2\) burst and produces no visible change in
\(\tilde{I}_{32}\). A larger \(R^0_{31}\) therefore strengthens
the radiative output of the non-inverted pathway without turning it
into an inversion-driven maser or superradiant channel.

In an astrophysical context, the fiducial \(c_{31}\) should be regarded
as an effective initial-coherence parameter rather than a directly
measured quantity. For specified populations and field coupling, an
observational upper limit on the \(3\rightarrow1\) intensity would
constrain \(c_{31}\). In general, however, it constrains the effective
coherence \(R^0_{31}=c_{31}\mathcal{R}_{31}^{\max}\), with
\(\mathcal{R}_{31}^{\max}=2\sqrt{n_3^0n_1^0}\), so \(c_{31}\) cannot be
determined independently of the initial populations.

Figure~\ref{fig:case_coherence}(b) shows the contrasting effect of
\(c_{32}\). A larger initial \(R^0_{32}\) provides a stronger
seed for the cooperative buildup of macroscopic polarization, causing
the \(3\rightarrow2\) superradiance burst to occur earlier and reach a higher peak
intensity. Thus, \(c_{31}\) primarily determines the visibility of the
leakage-channel emission, whereas \(c_{32}\) controls the initial
intensity, delay time, and strength of the collective response.


\subsection{Non-coherent Processes and Radiative Regimes}
\label{subsec:noncoherent_regimes}

The preceding subsections examined the roles of the initial population
distribution and coherence seeding in the absence of non-coherent
processes. We now allow relaxation and dephasing to act on finite
timescales, as expected in astrophysical gas, where collisions and other
environmental interactions can regulate the population reservoir and
limit the lifetime of phase coherence \citep{Rajabi2020}.

\begin{figure*}
    \centering
    \begin{minipage}{0.49\textwidth}
        \centering
        \textbf{(a)}\\[-1mm]
        \includegraphics[width=\linewidth]{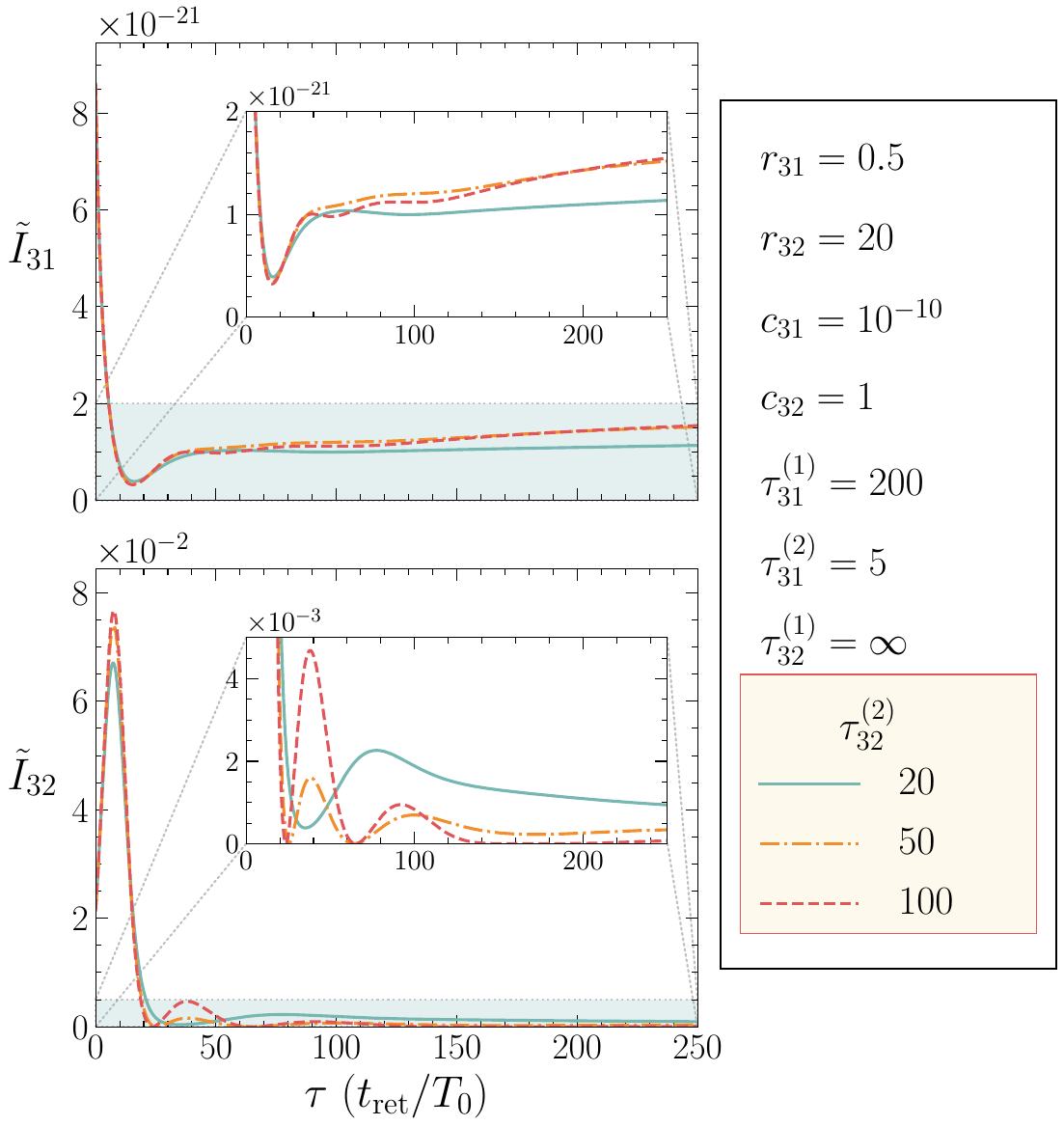}
    \end{minipage}
    \hfill
    \begin{minipage}{0.49\textwidth}
        \centering
        \textbf{(b)}\\[-1mm]
        \includegraphics[width=\linewidth]{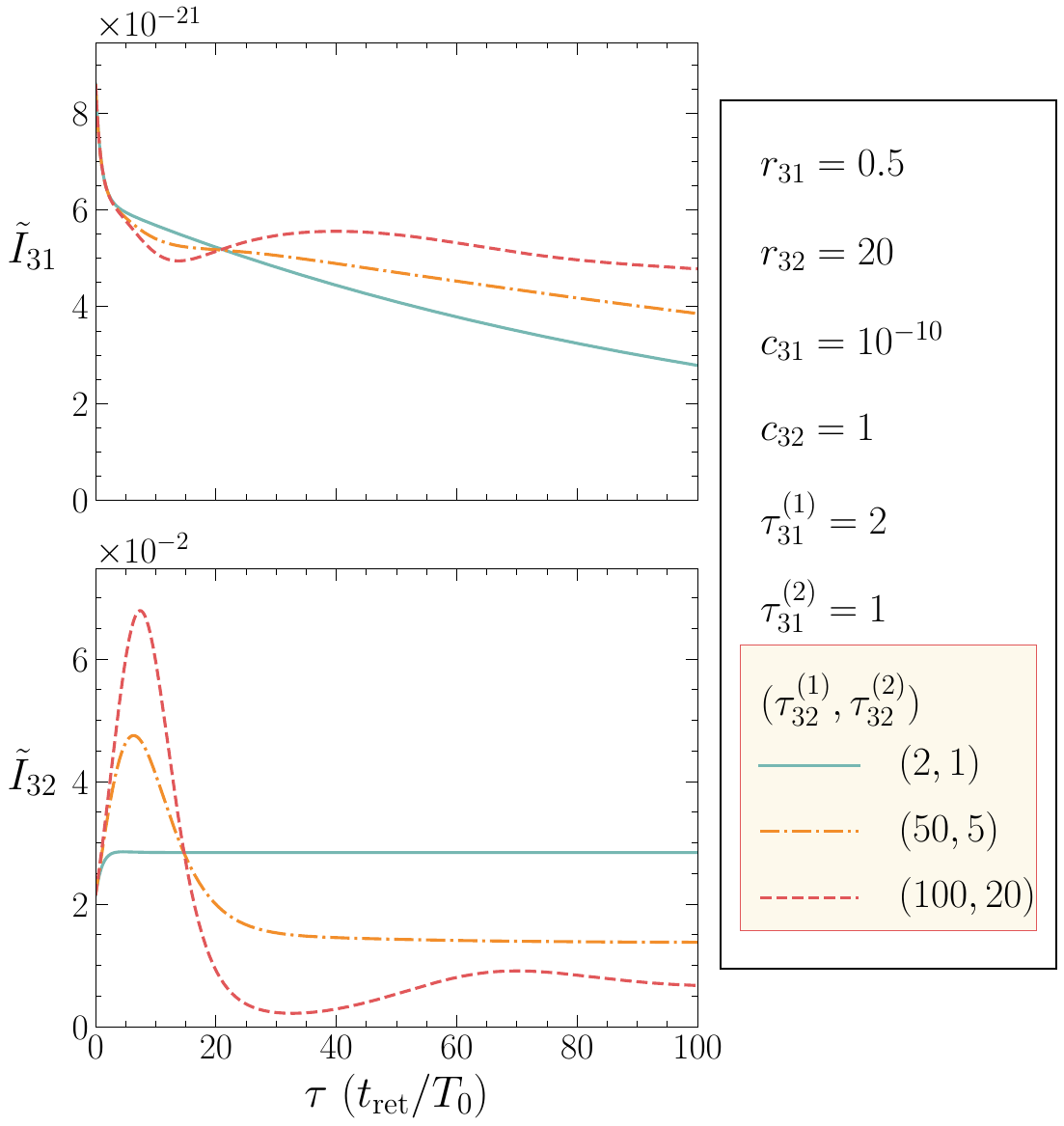}
    \end{minipage}
    \caption{Effect of non-coherent relaxation and dephasing in the
    inverted \(3\rightarrow2\) subsystem. The scaled output intensities,
    \(\tilde{I}_{31}(1,\tau)\) and \(\tilde{I}_{32}(1,\tau)\), are shown
    as functions of the scaled retarded time \(\tau\) at the output end
    of the sample, \(\xi=1\). In both panels, \(\eta=1\), the
    non-coherent timescales of the \(3\rightarrow1\) leakage pathway
    satisfy Eq.~\eqref{eq:timescales_inequality}, and all remaining
    parameters are listed in the boxes. Panel~(a) varies the
    \(3\rightarrow2\) dephasing time \(\tau^{(2)}_{32}\) while setting
    \(\tau^{(1)}_{32}=\infty\), so the \(3\rightarrow2\) inversion is
    not replenished by population relaxation. Decreasing
    \(\tau^{(2)}_{32}\) reduces the superradiant burst amplitude and
    damps the ringing, but the response remains transient and eventually
    decays. Panel~(b) varies the pair
    \((\tau^{(1)}_{32},\tau^{(2)}_{32})\): short non-coherent timescales
    interrupt synchronization and produce a smooth quasi-steady maser
    response (cyan line), whereas longer timescales allow a delayed superradiant
    burst and moderate ringing before the emission settles.}
    \label{fig:T32_scans}
\end{figure*}

Figure~\ref{fig:T32_scans}(a) examines the role of
\(3\rightarrow2\) dephasing by varying \(\tau_{32}^{(2)}\) while
setting \(\tau_{32}^{(1)}=\infty\). The \(3\rightarrow1\) relaxation
and dephasing timescales are held fixed at the fiducial values listed
in the figure. When \(\tau_{32}^{(2)}\) is long compared with the
timescale for collective polarization to develop, phase coherence
survives long enough for the molecular dipoles to synchronize,
producing a delayed superradiant burst followed by ringing. Shortening
\(\tau_{32}^{(2)}\) weakens the phase correlations before full
synchronization can occur, reducing the burst amplitude and damping
the ringing. Because \(\tau_{32}^{(1)}=\infty\), the population
difference is not restored toward the reference inversion
\(n_{32}^{0}\); the response therefore remains transient, and all
curves eventually decay.

Figure~\ref{fig:T32_scans}(b) shows the response when both
\(3\rightarrow2\) timescales are finite. When
\(\tau_{32}^{(1)}\) and \(\tau_{32}^{(2)}\) are short, rapid dephasing
suppresses the buildup of macroscopic coherence, while relaxation
restores \(n_{32}\) toward \(n_{32}^{0}\), causing the intensity to rise
smoothly toward a quasi-steady maser level, as shown by the cyan solid
curve. As the timescales increase, phase coherence survives long enough
for collective synchronization to develop, producing a transient
superradiant burst and ringing before the emission settles. Thus,
\(\tau_{32}^{(2)}\) determines whether macroscopic coherence can
develop, whereas a finite \(\tau_{32}^{(1)}\) replenishes the inversion
needed to sustain quasi-steady emission. The quasi-steady maser and
transient superradiant regimes of the effective two-level treatment
\citep{Rajabi2020} therefore remain present in the mixed three-level
system despite the leakage pathway.

The non-coherent timescales of the \(3\rightarrow1\) leakage pathway
play a secondary role for the fiducial parameters considered here.
Because this pathway is non-inverted and weakly seeded, dephasing mainly
damps the effective coherence that sources its weak field, while
population relaxation restores the population difference toward its
non-inverted reference value. These processes regulate the
leakage-channel signal but do not significantly alter the dominant
\(3\rightarrow2\) response, as further illustrated in
\hyperref[app:noncoherent_effects]{Appendix~B}.

\subsection{Application to the 6.7~GHz Methanol Flare in S255IR-NIRS3}
\label{subsec:S255}

\begin{figure}
    \centering
    \includegraphics[width=0.8\linewidth]{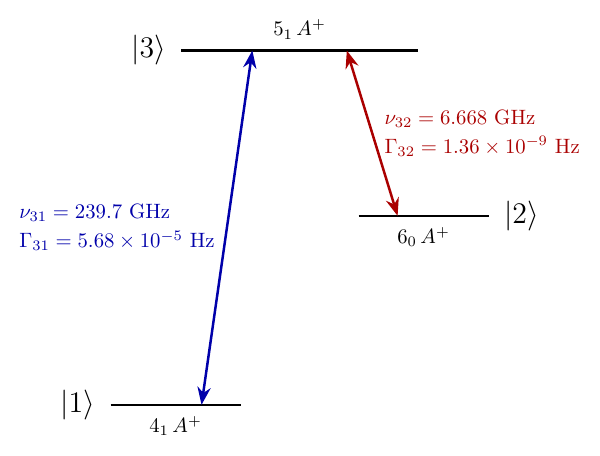}
    \caption{Methanol level structure used for the three-level MBE
    model of the 6.7~GHz flare in S255IR-NIRS3. The inverted
    \(3\rightarrow2\) transition corresponds to the 6.668~GHz
    \(5_{1}A^{+}\rightarrow6_{0}A^{+}\) methanol line, while the
    competing \(3\rightarrow1\) pathway corresponds to the fast
    239.7~GHz \(5_{1}A^{+}\rightarrow4_{1}A^{+}\) decay channel.
    The two transitions share the same \(5_{1}A^{+}\) upper level,
    but the spontaneous decay rate of the \(3\rightarrow1\) pathway is more than four orders of magnitude greater than that of the 6.668~GHz transition. In the \(J_{K}\) notation, \(J\) denotes the rotational angular momentum quantum number and \(K\) its projection along the molecular symmetry axis. The level spacings are schematic and not to scale.}
    \label{fig:methanol_scheme}
\end{figure}

We apply the three-level MBE framework to the 6.7~GHz methanol flare in S255IR-NIRS3. The observational data used here were originally obtained with the Toruń 32~m radio telescope and reported by \citet{Szymczak2018a,Szymczak2018b}. These data were previously modelled within a two-level MBE framework by \citet{Rajabi2019}. No new observational data were obtained for the present study.

S255IR-NIRS3 provides a concrete benchmark for the mixed configuration studied above. The observed 6.668~GHz methanol transition, commonly referred to as the 6.7~GHz line, shares its upper level with the much faster 239.7~GHz radiative leakage pathway, as shown in Fig.~\ref{fig:methanol_scheme}. We use this case to test whether the 6.7~GHz transition can retain its inversion and support a transient superradiant burst when the known fast decay channel is included
explicitly.

\begin{figure*}[t]
    \centering
    \includegraphics[width=0.5\textwidth]{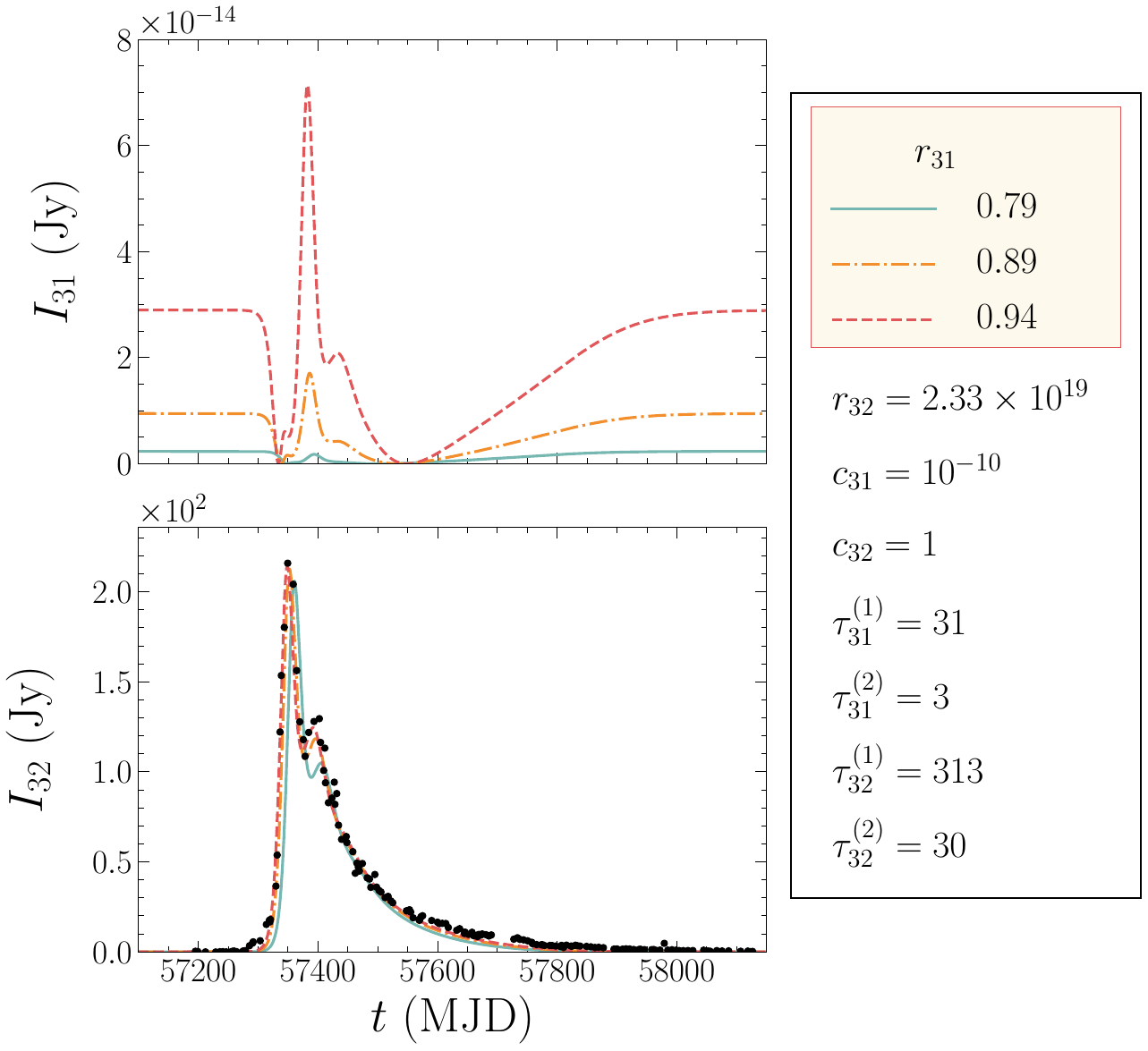}
    \caption{Three-level MBE modelling of the 6.7~GHz methanol flare in S255IR-NIRS3 for the \(v_{\mathrm{lsr}}=6.93~\mathrm{km\,s^{-1}}\) velocity component. The upper panel shows the intensity predicted for the competing 239.7~GHz \(3\rightarrow1\) leakage pathway, while the lower panel shows the corresponding 6.7~GHz \(3\rightarrow2\) emission together with the observational data obtained with the Toruń 32~m radio telescope and reported by \citet{Szymczak2018a,Szymczak2018b} (black points). The three curves correspond to different values of the initial population ratio \(r_{31}=n_3^0/n_1^0\) associated with the approximately thermalized \(3\rightarrow1\) subsystem, while the spectroscopic field-coupling ratio is fixed at \(\eta=27.51\). The same pump pulse is used for all three curves and is defined by Eq.~\eqref{eq:pump}, with dimensional amplitude \(\Lambda_{1}^{\mathrm{dim}}
    =2.6\times10^{-19}\,\mathrm{cm^{-3}\,s^{-1}}\) and characteristic duration \(T_{\mathrm{p}}=8.1\times10^{6}\,\mathrm{s}\),
    corresponding to \(\tau_{\mathrm{p}}\simeq155\). The remaining
    fixed model parameters are listed in the parameter box.}
    \label{fig:case_r1}
\end{figure*}

Consistent with the earlier two-level MBE modelling of the 6.7~GHz
flare by \citet{Rajabi2019}, we retain \(r_{32}=2.33\times10^{19}\) and \(c_{32}=1\) for the observed \(3\rightarrow2\) transition, while adding a weakly coherent leakage pathway with \(c_{31}=10^{-10}\). The non-coherent timescales are expressed in units of \(T_0\simeq5.23\times10^{4}\,\mathrm{s}\) and are chosen to be shorter for the \(3\rightarrow1\) pathway than for the inverted transition. The full parameter set is shown in the parameter box of Fig.~\ref{fig:case_r1}. These choices place S255IR-NIRS3 in the
strong-inversion limit of the mixed configuration, with the
\(3\rightarrow1\) transition treated as an approximately thermalized
leakage pathway.

Following \citet{Rajabi2019}, we model the pumping as a steady
background component supplemented by a transient enhancement associated with the infrared outburst of the central source
\citep{Caratti2017,Szymczak2018b}. The steady pump--loss balance is represented phenomenologically by the relaxation term in
Eq.~\eqref{eq:n32_dimless}, which drives \(n_{32}\) toward the reference inversion \(n_{32}^{0}\). The flare is initiated by adding the same pump pulse used in the earlier two-level fit as a source term on the right-hand side of the \(n_{32}\) equation:
\begin{equation}
\tilde{\Lambda}(\tau)=
\frac{\tilde{\Lambda}_{1}}
{\cosh^{2}\!\left[(\tau-\tau_{0})/\tau_{\mathrm{p}}\right]} .
\label{eq:pump}
\end{equation}
Here, \(\tilde{\Lambda}_{1}\), \(\tau_{\mathrm{p}}\), and \(\tau_{0}\)
denote the scaled pulse amplitude, duration, and temporal offset,
respectively. We adopt the dimensional pulse parameters from the
two-level model of \citet{Rajabi2019},
\(\Lambda_{1}^{\mathrm{dim}}=2.6\times10^{-19}\,
\mathrm{cm^{-3}\,s^{-1}}\) and
\(T_{\mathrm{p}}=8.1\times10^{6}\,\mathrm{s}\), which give
\(\tau_{\mathrm{p}}=T_{\mathrm{p}}/T_0\simeq155\).

Using spectroscopic data from the CDMS \citep{CDMS2001}, we adopt the transition dipole moments \(d_{31}\simeq0.594~\mathrm{D}\) and
\(d_{32}\simeq0.679~\mathrm{D}\). Together with the transition
frequencies shown in Fig.~\ref{fig:methanol_scheme}, these values yield \(\eta=27.51\) through Eq.~\eqref{eq:eta}.

Figure~\ref{fig:case_r1} shows that the population ratio \(r_{31}\)
affects the morphology of the 6.7~GHz flare. A larger \(r_{31}\) provides a greater shared upper-level population reservoir, thereby modifying the timing and amplitude of the \(3\rightarrow2\) burst, as also seen in Fig.~\ref{fig:case_population}. If the \(3\rightarrow1\) subsystem is thermalized, \(r_{31}\) can be related to its excitation temperature through the Boltzmann population ratio. Fitting the flare profile may therefore constrain \(r_{31}\) and, when combined with an appropriate excitation model, place constraints on the temperature, density, and radiative environment of the flaring gas.

For the fiducial small value of \(c_{31}\), the fast 239.7~GHz channel does not eliminate the superradiant response of the 6.7~GHz transition. The transient-superradiance interpretation obtained with the earlier two-level model remains compatible with a three-level treatment that explicitly includes the leakage pathway. In this regime, the \(3\rightarrow1\) output remains far weaker than the \(3\rightarrow2\) emission.

Scans over \(c_{31}\) further show that strengthening the effective coherence of the leakage channel primarily enhances the weak 239.7~GHz signal without significantly altering the 6.7~GHz flare profile, as also seen in Fig.~\ref{fig:case_coherence}. Because the spectroscopic coupling is fixed, an observational upper limit on the 239.7~GHz leakage-channel emission would translate into an upper limit on the effective \(3\rightarrow1\) coherence.

The S255IR-NIRS3 calculation is a representative strong-inversion
benchmark rather than an exhaustive survey of shared-upper-level
masers. The same framework applies to other methanol or OH transitions
with fast shared-upper-level decay channels once the spectroscopic
parameters, population ratios, coherence fractions, and non-coherent
timescales are replaced with values appropriate to those systems.

\section{Conclusions}
\label{sec:conclusion}

We have developed a \(\Lambda\)-type three-level Maxwell--Bloch model
to test whether maser amplification or transient superradiance can
survive when an inverted transition shares its upper level with a much
faster radiative decay pathway. Although the equations can describe
different initial population configurations, this paper focuses on the
mixed configuration relevant to shared-upper-level astronomical masers:
the \(3\rightarrow1\) pathway is non-inverted and acts as a leakage
channel, while the \(3\rightarrow2\) transition is inverted and can
support maser amplification or superradiance.

Our main result is that a large spontaneous decay rate in the competing
pathway does not, by itself, suppress maser amplification or
superradiant emission from the inverted transition. The outcome depends
on the population ratios, which set the shared upper-level reservoir,
the \(3\rightarrow2\) inversion, and the maximum available coherence.
Thus, a larger inversion does not necessarily produce a stronger
superradiant burst if it is accompanied by a weaker initial coherence
seed. The effective \(3\rightarrow1\) coherence controls the weak
leakage-channel output, whereas the initial \(3\rightarrow2\)
coherence controls the delay and strength of the collective response.
Finally, the \(3\rightarrow2\) relaxation and dephasing timescales
determine whether the system exhibits quasi-steady maser amplification
or transient superradiance, while the corresponding leakage-channel
timescales play a secondary role in the cases considered here.

We applied the model to the 6.7~GHz methanol flare in S255IR-NIRS3,
where the observed \(5_{1}A^{+}\rightarrow6_{0}A^{+}\) transition
shares its upper level with the much faster 239.7~GHz
\(5_{1}A^{+}\rightarrow4_{1}A^{+}\) decay pathway. Including this
channel explicitly does not eliminate the transient superradiant
response of the 6.7~GHz transition. Although it removes population from
the shared upper level, its radiative output remains weak for small
effective \(3\rightarrow1\) coherence. The dependence of the flare
profile on \(r_{31}\), together with observational limits on the
239.7~GHz emission, may constrain the shared upper-level population
reservoir and the effective coherence of the leakage pathway.

More generally, fast shared-upper-level decay should not be regarded as
a fundamental objection to maser amplification or superradiance in
multilevel molecular systems. It suppresses the observed response only
if it removes sufficient population to destroy the inversion or if
relaxation and dephasing prevent the required coherence dynamics from
developing. Otherwise, the fast transition can act primarily as a
population-leakage channel while the inverted transition remains the
dominant emitter. The relevant criterion is therefore the dynamical
competition among population leakage, inversion replenishment, and
coherence evolution, rather than the spontaneous decay rate of the
competing transition alone.


\begin{acknowledgments}
F.R.'s research is supported by the NSERC Discovery Grant RGPIN-2024-06346. F.R. is grateful for the hospitality of Perimeter Institute where part of this work was carried out. Research at Perimeter Institute is supported in part by the Government of Canada through the Department of Innovation, Science and Economic Development and by the Province of Ontario through the Ministry of Colleges and Universities. This work was supported by a grant from the Simons Foundation (1034867, Dittrich). During the preparation of this work, the authors used ChatGPT and Claude Opus 5.5 to improve readability through grammar and language refinement. The authors reviewed and edited the output as needed and take full responsibility for the content of the publication.
\end{acknowledgments}


\appendix

\section{Envelope Conventions and Maxwell--Bloch Equations}
\label{app:three_level_mbes}

We summarize the assumptions and envelope conventions used to obtain the
scaled equations in Sec.~\ref{sec:modeloverview}. The derivation follows
the standard density-matrix and Maxwell-equation treatment of resonant
multilevel media under the rotating-wave and slowly varying envelope
approximations \citep{Gross1982,Benedict1996}. We consider a
\(\Lambda\)-type system with level energies
\(\hbar\omega_1<\hbar\omega_2<\hbar\omega_3\), so that
\(\omega_{3j}=\omega_3-\omega_j\) and \(\omega_{31}>\omega_{32}\). The two
allowed transitions \(3\rightarrow1\) and \(3\rightarrow2\) are
coupled to classical fields at their carrier angular frequencies
\(\omega_{31}\) and \(\omega_{32}\), respectively. Detunings are set to zero throughout this work.

The state of the three-level molecules at position \(z\) is described
by the single-molecule density matrix
\begin{equation}
\hat{\rho}(z,t)=\sum_{i,j=1}^{3}
\rho_{ij}(z,t)\lvert i\rangle\langle j\rvert ,
\nonumber
\end{equation}
where \(\rho_{ij}\) are dimensionless density-matrix elements. The
diagonal elements \(n_i\equiv\rho_{ii}\) represent the fractional level
populations, where initially \(\Sigma_i n_i=1\).
The off-diagonal elements \(\rho_{ij}\) represent the corresponding quantum coherences. 

The density matrix evolves according to
\begin{equation}
\frac{\partial\hat{\rho}}{\partial t}
=
-\frac{i}{\hbar}\big[\hat{H},\hat{\rho}\big]
+\text{relaxation terms},
\label{eq:app_liouville}
\end{equation}
with the Hamiltonian
\begin{equation}
\hat{H}=\sum_{i=1}^{3}\hbar\omega_i\lvert i\rangle\langle i\rvert
-\hat{\mathbf{d}}\cdot\boldsymbol{\mathcal{E}}(z,t),
\nonumber
\end{equation}
where \(\hat{\mathbf{d}}\) is the electric-dipole operator and
\(\boldsymbol{\mathcal{E}}\) is the total electric field.

The total field is the sum of the two transition fields,
\(\boldsymbol{\mathcal{E}}=\boldsymbol{\mathcal{E}}_{31}+\boldsymbol{\mathcal{E}}_{32}\),
which are taken to propagate in the positive \(z\)-direction as plane
waves with fixed transverse polarization unit vectors
\(\hat{\mathbf{e}}_{3j}\). The real electric fields are written in terms
of slowly varying complex envelopes as
\begin{equation}
\boldsymbol{\mathcal{E}}_{3j}(z,t)
=
\frac{1}{2}\hat{\mathbf{e}}_{3j}E_{3j}(z,t)\,
e^{-i(\omega_{3j}t-k_{3j}z)}
+\mathrm{c.c.},
\quad j=1,2,
\label{eq:app_field}
\end{equation}
where \(k_{3j}=\omega_{3j}/c\). Here, \(E_{31}\) is the slowly varying
field envelope on the \(3\rightarrow1\) transition, while \(E_{32}\)
is the corresponding envelope on the \(3\rightarrow2\) transition.
The field on each allowed transition couples
through the effective transition-dipole component
\begin{equation}
d_{3j}\equiv\hat{\mathbf{e}}_{3j}\cdot\mathbf{d}_{3j},
\qquad
\mathbf{d}_{3j}=\langle 3\rvert\hat{\mathbf{d}}\lvert j\rangle .
\nonumber
\end{equation}

The coherences on the two dipole-allowed transitions are written in
terms of slowly varying amplitudes as
\begin{equation}
\rho_{3j}(z,t)
=
\frac{1}{2}R_{3j}(z,t)\,
e^{-i(\omega_{3j}t-k_{3j}z)},
\qquad j=1,2.
\label{eq:app_rho3j}
\end{equation}
Although no field is applied directly to the dipole-forbidden
\(2\leftrightarrow1\) transition, the lower-state coherence can be
generated indirectly through the two allowed transitions. We therefore
write
\begin{equation}
\rho_{21}(z,t)
=
\frac{1}{2}R_{21}(z,t)\,
e^{-i[(\omega_{31}-\omega_{32})t-(k_{31}-k_{32})z]} .
\label{eq:app_rho21}
\end{equation}
With these conventions, \(R_{31}\), \(R_{32}\), and \(R_{21}\) are slowly
varying coherence amplitudes, not macroscopic polarizations.
We describe the populations through the population differences
\(n_{3j}=n_3-n_j\), with \(j=1,2\), where positive \(n_{3j}\) corresponds to inversion on the
\(3\rightarrow j\) transition.

Substituting Eqs.~\eqref{eq:app_field}--\eqref{eq:app_rho21} into
Eq.~\eqref{eq:app_liouville}, applying the rotating-wave and slowly
varying envelope approximations, and transforming to the retarded time
\(t_{\rm ret}=t-z/c\), the matter equations, including phenomenological
population relaxation and coherence dephasing, are
\begin{subequations}
\label{eq:app_matter}
\begin{align}
\frac{\partial n_{31}}{\partial t_{\rm ret}}
&=
\frac{1}{\hbar}\,
\operatorname{Im}\!\left(
E_{31}^{*}d_{31}^{*}R_{31}
+
\frac{1}{2}E_{32}^{*}d_{32}^{*}R_{32}
\right)
-
\frac{n_{31}-n_{31}^{0}}{T^{(1)}_{31}},
\label{eq:app_n31}\\[2mm]
\frac{\partial n_{32}}{\partial t_{\rm ret}}
&=
\frac{1}{\hbar}\,
\operatorname{Im}\!\left(
\frac{1}{2}E_{31}^{*}d_{31}^{*}R_{31}
+
E_{32}^{*}d_{32}^{*}R_{32}
\right)
-
\frac{n_{32}-n_{32}^{0}}{T^{(1)}_{32}},
\label{eq:app_n32}\\[2mm]
\frac{\partial R_{31}}{\partial t_{\rm ret}}
&=
-\frac{i}{\hbar}E_{31}d_{31}n_{31}
+
\frac{i}{2\hbar}E_{32}d_{32}R_{21}
-
\frac{R_{31}-R_{31}^{0}}{T^{(2)}_{31}},
\label{eq:app_R31}\\[2mm]
\frac{\partial R_{32}}{\partial t_{\rm ret}}
&=
-\frac{i}{\hbar}E_{32}d_{32}n_{32}
+
\frac{i}{2\hbar}E_{31}d_{31}R_{21}^{*}
-
\frac{R_{32}-R_{32}^{0}}{T^{(2)}_{32}},
\label{eq:app_R32}\\[2mm]
\frac{\partial R_{21}}{\partial t_{\rm ret}}
&=
\frac{i}{2\hbar}
\left(
E_{32}^{*}d_{32}^{*}R_{31}
-
E_{31}d_{31}R_{32}^{*}
\right)
-
\frac{R_{21}-R_{21}^{0}}{T^{(2)}_{21}} .
\label{eq:app_R21}
\end{align}
\end{subequations}
Here, \(\operatorname{Im}(\cdot)\) denotes the imaginary part of the entire
expression enclosed in parentheses. The timescales \(T_{3j}^{(1)}\)
describe relaxation of the population differences \(n_{3j}\), while
\(T_{ij}^{(2)}\) describe dephasing of the coherences \(R_{ij}\). The
superscript ``0'' denotes the reference value toward which the
corresponding relaxation or dephasing term drives the variable.

The fields are driven by the resonant macroscopic polarization, which
contains only the dipole-allowed coherences,
\begin{equation}
\boldsymbol{\mathcal{P}}(z,t)
=
\frac{n_0}{2}\sum_{j=1}^{2}
\mathbf{d}_{3j}^{*}R_{3j}\,
e^{-i(\omega_{3j}t-k_{3j}z)}
+\mathrm{c.c.},
\label{eq:app_polarization}
\end{equation}
where \(n_0\) is the total initial number density of the three-level
molecules. Inserting Eqs.~\eqref{eq:app_field} and \eqref{eq:app_polarization}
into the one-dimensional wave equation
\begin{equation}
\left(\frac{\partial^2}{\partial z^2}
-\frac{1}{c^2}\frac{\partial^2}{\partial t^2}\right)
\boldsymbol{\mathcal{E}}
=
\mu_0\frac{\partial^2\boldsymbol{\mathcal{P}}}{\partial t^2},
\label{eq:app_wave}
\end{equation}
projecting onto \(\hat{\mathbf{e}}_{3j}^{*}\), keeping the resonant terms,
and applying the same approximations, the field equations for the slowly
varying envelopes, evaluated at fixed retarded time, are
\begin{equation}
\frac{\partial E_{3j}}{\partial z}
=
i\frac{n_0k_{3j}d_{3j}^{*}}{2\epsilon_0}R_{3j},
\qquad j=1,2.
\label{eq:app_E3j}
\end{equation}

Equations~\eqref{eq:app_matter} and \eqref{eq:app_E3j} are the unscaled
Maxwell--Bloch equations. Introducing the scaled variables of
Sec.~\ref{sec:modeloverview} gives
Eqs.~\eqref{eq:matter_dimless} and \eqref{eq:E3j_dimless}.
\section{Additional Non-coherent Timescale Effects in the Leakage Channel}
\label{app:noncoherent_effects}

In Sec.~\ref{subsec:noncoherent_regimes}, we showed that the
transition between quasi-steady maser emission and transient
superradiance is controlled primarily by the non-coherent timescales of
the inverted \(3\rightarrow2\) subsystem. Here we provide supporting
scans for the non-inverted \(3\rightarrow1\) leakage pathway, using the
fiducial weak coherence fraction \(c_{31}=10^{-10}\). These calculations
test whether changing the relaxation or dephasing time of the fast
pathway significantly modifies the dominant \(3\rightarrow2\) response.

\begin{figure*}[t]
    \centering
    \begin{minipage}{0.49\textwidth}
        \centering
        \textbf{(a)}\\[-1mm]
        \includegraphics[width=\linewidth]{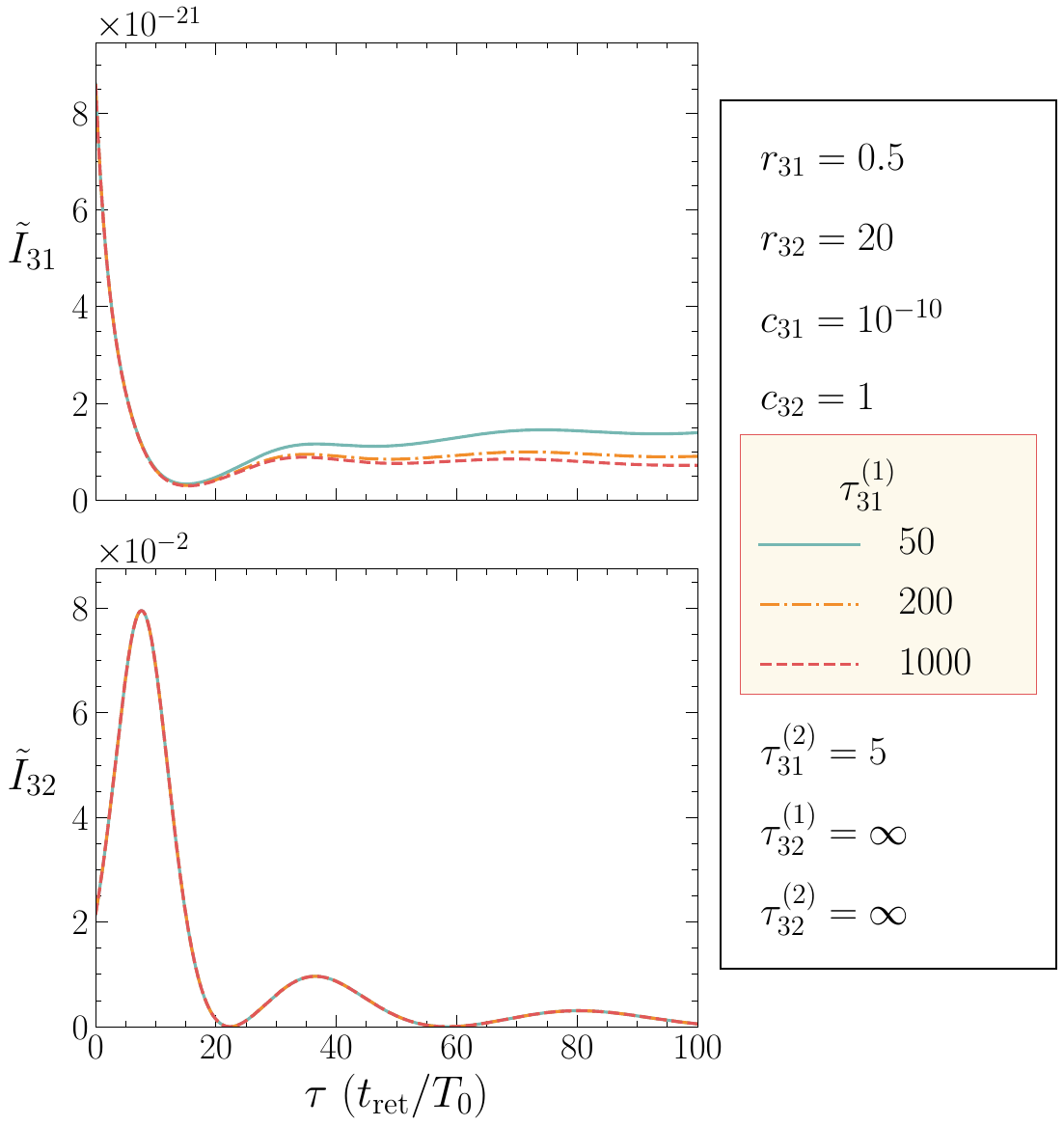}
    \end{minipage}
    \hfill
    \begin{minipage}{0.49\textwidth}
        \centering
        \textbf{(b)}\\[-1mm]
        \includegraphics[width=\linewidth]{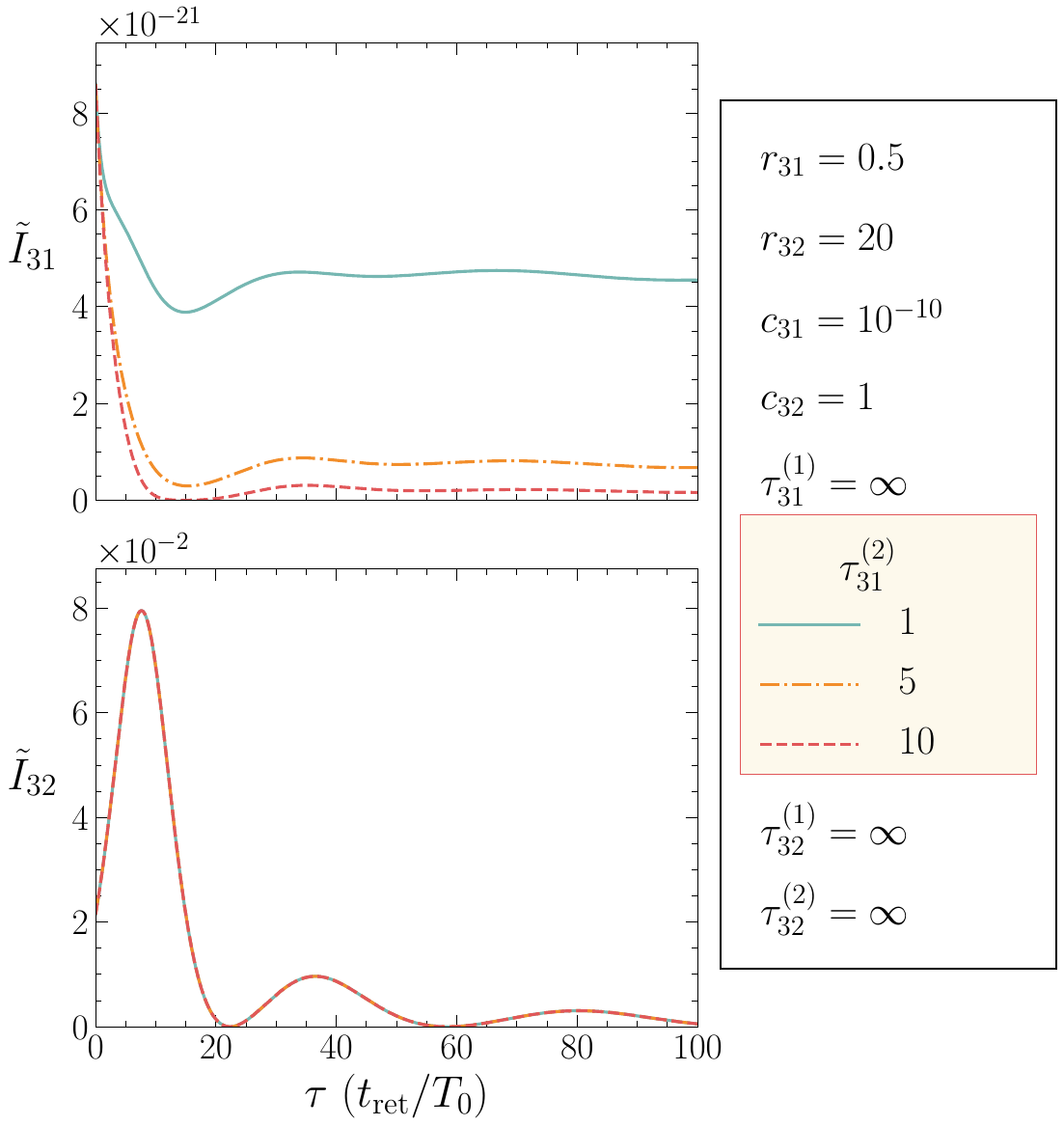}
    \end{minipage}
\caption{Effect of non-coherent relaxation and dephasing in the
non-inverted \(3\rightarrow1\) leakage pathway. The scaled output
intensities, \(\tilde I_{31}(1,\tau)\) and
\(\tilde I_{32}(1,\tau)\), are shown as functions of the scaled
retarded time \(\tau\) at the output end of the sample, \(\xi=1\).
Panel~(a) varies the population-relaxation time
\(\tau_{31}^{(1)}\), while panel~(b) varies the coherence-dephasing
time \(\tau_{31}^{(2)}\). In both panels, \(\eta=1\), so the two
radiative pathways have equal scaled field-coupling strength; all
other parameters are fixed as listed in the figure boxes.}
\label{fig:appendix_T31_scans}
\end{figure*}

Figure~\ref{fig:appendix_T31_scans}(a) shows that changing the
population-relaxation time of the \(3\rightarrow1\) pathway mainly
affects the weak leakage-channel output. Because this transition is
non-inverted, finite relaxation drives \(n_{31}\) back toward its
non-inverted reference value rather than supplying gain. The resulting
changes are therefore confined primarily to \(\tilde I_{31}\), while
the collective \(3\rightarrow2\) emission remains essentially unchanged
for the parameters shown.

Figure~\ref{fig:appendix_T31_scans}(b) shows the corresponding effect
of varying the \(3\rightarrow1\) dephasing time. Shorter
\(\tau_{31}^{(2)}\) damps the effective coherence
\(\mathcal R_{31}\) that sources the leakage-channel field, reducing
transient structure in \(\tilde I_{31}\). As in the
population-relaxation scan, the dominant \(3\rightarrow2\) emission is
insensitive to these changes for the adopted small \(c_{31}\), for
which the leakage pathway remains weakly radiative.

These supplementary scans justify treating the \(3\rightarrow1\)
non-coherent timescales as secondary in the fiducial mixed-regime
calculations with small \(c_{31}\). They regulate the weak
leakage-channel output but do not determine whether the dominant
\(3\rightarrow2\) response is maser-like or superradiant; that
transition is controlled primarily by the relaxation and dephasing
timescales of the inverted subsystem.

\bibliographystyle{apsrev4-2}
\bibliography{refs}

\end{document}